\documentclass[showpacs,aps,graphicx]{revtex4}

\usepackage{graphicx}

\usepackage{amsmath}

\begin{document}

\title{Scalable photonic quantum computing assisted by  quantum-dot spin
in double-sided optical microcavity\footnote{Published in Opt.
Express \textbf{21}, 17671-17685 (2013)}}
\author{Hai-Rui Wei and Fu-Guo Deng\footnote{Corresponding author: fgdeng@bnu.edu.cn} }
\address{ Department of Physics, Applied Optics Beijing Area Major Laboratory,
Beijing Normal University, Beijing 100875, China}
\date{\today }

\begin{abstract}

We investigate the possibility of achieving scalable photonic
quantum computing by the giant optical circular birefringence
induced by a quantum-dot spin in a double-sided optical microcavity
as a result of cavity quantum electrodynamics. We construct a
deterministic controlled-not  gate on two photonic qubits by two
single-photon input-output processes and the readout on an
electron-medium spin confined in an optical resonant microcavity.
This idea could be applied to multi-qubit gates on photonic qubits
and we give the quantum circuit for a three-photon Toffoli gate.
High fidelities and high efficiencies could be achieved when the
side leakage to the cavity loss rate is low. It is worth  pointing
out that our devices work in both the strong   and  the weak
coupling regimes.

\end{abstract}

 \maketitle


\section{Introduction}\label{sec1}

Quantum information processing requires the precise control and
manipulation of quantum states. It has been proven that single-qubit
operations and two-qubit entangling gates are sufficient for
universal quantum computing \cite{universal}. One of the typical
two-qubit gates is the controlled-not (CNOT) gate. The
implementation of a photonic CNOT gate is one of the main directions
as photons can be transmitted fast and reliably over a long distance
in  a less-loss optical fiber or a free space (low decoherence) and
a single-qubit manipulation can be accomplished easily. Although
they are perfect information carriers in quantum communication
purpose, photons seem to be less suitable for quantum computing as
they lack the sufficient strong interaction between each other.
Surprisingly, Knill, Laflamme, and Milburn  \cite{KLM} demonstrated
that using only single-photon sources, single-photon detectors, and
linear optical elements, a CNOT gate on photonic qubits with a
maximal success probability of 3/4 could be created with the use of
polynomial resources. Subsequently, some improved works in theory
\cite{theor1,theor2,theor3,theor4} and in experiment
\cite{experi1,experi2,experi3} have been accomplished.
Unfortunately,  Pittman, Jacobs, and  Franson \cite{theor1} showed
that even with the help of maximally entangled two-photon pairs, the
success probability of a CNOT gate can only be boosted to $1/4$ and
it is still far lower than the maximum $3/4$.

Toffoli gate is a fundamental quantum gate for three-qubit systems.
It has been demonstrated that an arbitrary multi-qubit  gate can be
decomposed into a sequence of Toffoli  and Hadamard gates
\cite{Toffoli}. Linear optical three-qubit Toffoli gate with an
optimal success probability of 0.75\% is proposed by
Fiur\'{a}\v{s}ek \cite{Linear-optics}.  In 2009, Shende and Markov
\cite{optimal} proved that 6 CNOT gates are required for the
synthesis of  a Toffoli gate in the best case, which increases the
difficulty to implement the  Toffoli gate.

To avoid the probabilistic quantum computing with linear optics, a
near deterministic CNOT gate based on weak cross-Kerr nonlinearities
has been proposed \cite{kerr1,kerr2} and it provides a way for
deterministic quantum computation in principle
\cite{comput1,comput2}.  Based on cross-Kerr nonlinearity, Lin and
He proposed a scheme for a three-qubit Toffoli gate
\cite{Toffoli-Kerr}, and the scheme contains 4 two-qubit gates and 2
additional photonic qubits are required. A giant Kerr nonlinearity
is still a challenge with current technology, even with
electromagnetically induced transparency \cite{EIT1}, as the
initially achieved phase shift at the single-photon level is only in
the order of $10^{-5}$ \cite{EIT2}. Moreover, it is currently not
clear whether these nonlinearities are sufficient for the natural
implementation of single-photon qubit gates.

Previous works showed that the spin of a singly charged electron
confined in a quantum dot (QD) \cite{QD1,QD2,QD3,QD4,QD5,QD6} can be
used for storing and processing quantum information, due to the long
electron-spin coherence time ($\sim\mu s$)
\cite{cohertime1,cohertime2,cohertime3,cohertime4,cohertime5,cohertime6}
using spin-echo techniques. The QD spin state can be initiated by
means of optical  pumping or optical cooling \cite{eigen1,eigen2},
and then by performing single-spin rotations
\cite{cohertime1,cohertime2} or via a spin-flip Raman transition
\cite{eigen1}, the spin superposition state can be obtained.
Ultrafast optical coherent control on QD spins has been demonstrated
\cite{manipulating1,manipulating2,manipulating3}. The system based
on QD has received increasing attention because of the comparative
easiness of incorporating a QD into a solid-state cavity, which
could facilitate the deterministic transfer of quantum information
between the photonic qubit and the spin qubit, and this
transformation extends the potential applications of quantum
information and quantum communication. A spin-QD-cavity unit, e.g.,
an excess electron confined in a self-assembled In(Ga)As QD or a
GaAs interface QD inside an optical resonant microcavity was
proposed by Hu \emph{et al.} \cite{Hu1,Hu2}. This unit is the key
element for constructing universal hybrid quantum gates, entangled
state analyzer, teleportation, deterministic photonic hyper-CNOT
gates, quantum repeaters, entanglement purification, and creating
entangled states and hyperentangled Bell states
\cite{Hu1,Hu2,Hu3,Hu4,Appli1,weinot,aa0,aa1,aa2,aa3,aa4,aa5}. In
this unit, the QD spin represents the qubit and promises a scalable
quantum computing, and the QD spin manipulation is well developed
using the pulsed magnetic-resonance technique.

In this article, we investigate the possibility of achieving
scalable photonic quantum computing, assisted by a QD spin in a
double-sided optical microcavity. By the giant optical circular
birefringence for the  right-circularly and the left-circularly
lights induced by a QD spin in a microcavity as a result of cavity
quantum electrodynamics (QED), we construct a  CNOT gate  and a
Toffoli gate for photonic qubits. Both these two quantum gates are
based on some single-photon input-output processes and the readout
on an electron-medium spin. They work in a deterministic way. The
CNOT and the Toffoli gates discussed in \cite{weinot,Appli1} are
encoded on a control photon polarization qubit (or an electron-spin
qubit) and target electron spin qubits (photon polarization qubits).
Both the present quantum gates are encoded on the polarization of
single photonic qubits, and they have several advantages, including
scalable, low decoherence, deterministic, and suitable for
information transmission. The electron confined in the cavity only
plays a role of the medium. We analyze the experimental feasibility
of these two quantum gates, concluding that our proposals can be
implemented with current technology.

\section{Deterministic photonic two-qubit CNOT gate without additional  photonic  qubits}\label{sec2}

\begin{figure}
\centering
\includegraphics[width=6.4 cm,angle=0]{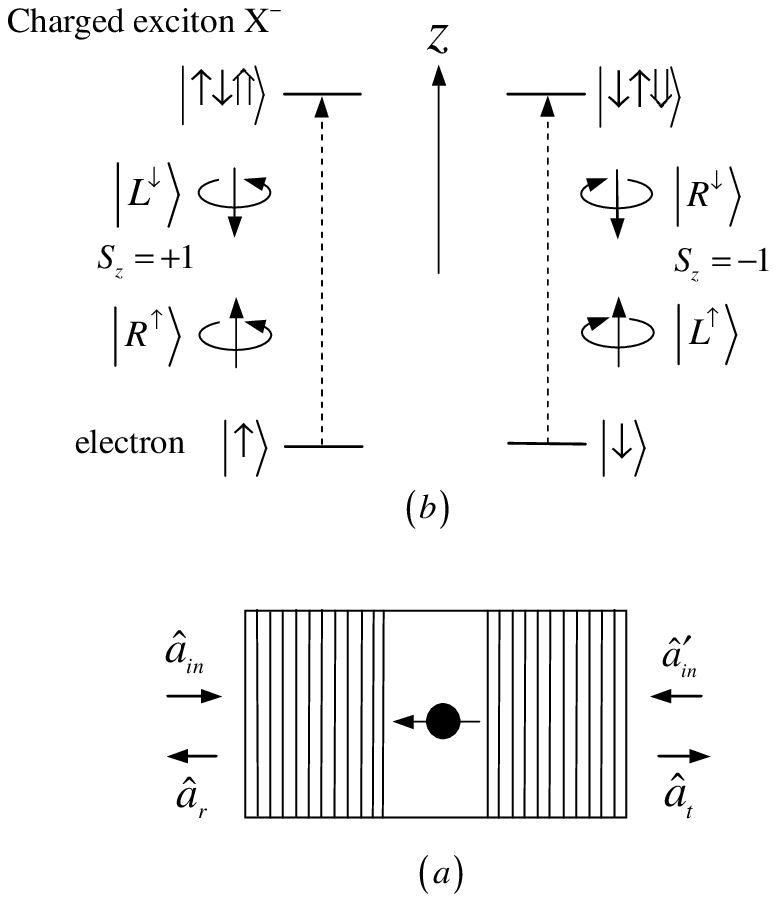}
\caption{ (a) Schematic diagram for a spin-QD-double-side-cavity
unit. (b) Energy levels and optical property of a negatively charged
exciton $X^-$ in a GaAs/InAs QD or GaAs interface QD confined in an
optical resonant microcavity with two partially reflective mirrors.}
\label{Fig1}
\end{figure}

Let us consider a singly charged electron self-assembled GaAs/InAs
QD or GaAs interface QD confined in an optical resonant microcavity
with two partially reflective mirrors \cite{Hu2,Appli1} (called it a
spin-QD-double-side-cavity unit below, see Fig. \ref{Fig1}(a)). When
an excess electron is injected into the QD, optical excitation can
create a trion $X^-$ (also called negatively charged exciton) which
consists of two electrons and a hole \cite{exciton1}. The optical
properties of the unit are determined by $X^-$, and there are only
two dipole transitions which lead to large differences in the phase
or the amplitude of the reflection and the transmission coefficients
between two circular polarizations of photons, one involving a
$s_z=+1$ photon and the other involving a $s_z=-1$ photon (see Fig.
\ref{Fig1}(b)) \cite{exciton2} due to Pauli's exclusion principle.
In the following, we consider the case that a dipole is resonant
with the cavity mode and is probed with a resonant light. If the
injected photon  couples to the dipole, that is, the photon can be
absorbed to create a $X^-$, the cavity is reflective, and both the
polarization and the propagation direction of the photon will be
flipped. Otherwise, the cavity is transmissive and the photon will
acquire a $\pi$ mod $2\pi$ phase shift relative to the reflected
photon. The interaction between the input photons with $s_z = \pm1$
and the electron spin  in the cavity is described as follows
\cite{Hu2,Appli1,weinot}:
\begin{equation}       \label{eq1}
\begin{split}
|R^\uparrow\uparrow\rangle &\;\;\;\;\rightarrow\;\;\;\;
|L^\downarrow\uparrow\rangle,
\;\;\;\;\;\;\;\;\;\;\;\;\;\;\;\;\;\;\;\;\;\,     |L^\downarrow\uparrow\rangle \;\;\;\;\rightarrow\; \;\;\; |R^\uparrow\uparrow\rangle,  \\
|R^\downarrow\downarrow\rangle &\;\;\;\;\rightarrow\;\;\;\;
|L^\uparrow\downarrow\rangle,
\;\;\;\;\;\;\;\;\;\;\;\;\;\;\;\;\;\;\;\;\;\,  |L^\uparrow\downarrow\rangle \;\;\;\;\rightarrow\;\;\;\; |R^\downarrow\downarrow\rangle, \\
|R^\downarrow\uparrow\rangle &\;\;\;\;\rightarrow\;\;\;\;
-|R^\downarrow\uparrow\rangle,
\;\;\;\;\;\;\;\;\;\;\;\;\;\;\;\;\;\;        |L^\uparrow\uparrow\rangle \;\;\;\;\rightarrow\;\;\;\; -|L^\uparrow\uparrow\rangle, \\
|R^\uparrow\downarrow\rangle &\;\;\;\;\rightarrow\;\;\;\;
-|R^\uparrow\downarrow\rangle, \;\;\;\;\;\;\;\;\;\;\;\;\;\;\;\;\;\;
|L^\downarrow\downarrow\rangle \;\;\;\;\rightarrow\;\;\;\;
-|L^\downarrow\downarrow\rangle.
\end{split}
\end{equation}
Here $|\uparrow\rangle$ and $|\downarrow\rangle$ represent the spin
states $|+\frac{1}{2}\rangle$ and $|-\frac{1}{2}\rangle$ of the
excess electron, respectively. $|\Uparrow\rangle$ and
$|\Downarrow\rangle$ represent the spin states
$|+\frac{3}{2}\rangle$ and $|-\frac{3}{2}\rangle$ of  the
heavy-hole, respectively. The spin quantization axis for angular
momentum is along the normal direction of the cavity, that is, the
$z$ axis. The right-circularly-polarized photon and the
left-circularly-polarized photon marked by $|R\rangle$ and
$|L\rangle$, respectively, and the subscript $\uparrow$
($\downarrow$) indicates their propagation direction along (against)
the $z$ axis. Such a unit can act as an entanglement beam splitter
(EBS) which can splits directly an initial product state of a
photon-spin system into two entangled states via the transmission
and reflection in a deterministic way. It is worth pointing out that
a polarization-degenerate cavity mode is required  to transfer the
polarization of the photon to the spin of an atom-like system, or
vice the versa in several quantum information applications, and it
is also necessary in our schemes. Excellent progress has been made
on the unpolarized micropillar cavities
\cite{unpolarized-pillar1,unpolarized-pillar2,unpolarized-pillar3}
and the H1 photonic crystal cavities
\cite{unpolarized-photon1,unpolarized-photon2}.

In a single electron charged QD system, the spontaneous spin flip
Raman scattering transitions,
$|\downarrow\uparrow\Downarrow\rangle\rightarrow|\uparrow\rangle$
and
$|\uparrow\downarrow\Uparrow\rangle\rightarrow|\downarrow\rangle$,
are ideally dark \cite{dark}, nevertheless, in a realistic QD
system, there is $\Gamma\gg\gamma\neq0$ due to the inherent  hole
mixing or the factor that a in-plane magnetic field is not parallel
to the $z$ axis. Here, $\Gamma$ is the allowed transition and
$\gamma$ is the forbidden transition \cite{eigen1}. The strong
hyperfine interaction between the excess electron and the QD nuclear
spin ensemble leads to a spin flip
($|\downarrow\rangle\leftrightarrow|\uparrow\rangle$) at the rate
$\xi_{\uparrow\downarrow}$. $\xi_{\uparrow\downarrow}$ is strongly
suppressed even under a weak magnetic field $B$
\cite{suppress1,suppress2}. When $B=0$, the optical transitions do
not alter the spin state occupancies\cite{eigen1}. The energy
splitting, which results in a net spin flip, does not occur for
charged excitons due to the quenched exchange interaction
\cite{occur1,occur2}, and   our schemes are immune to energy
splitting in principle.

In the following, we will utilize this EBS to construct a photonic
two-qubit CNOT gate.

\begin{figure}
\centering
\includegraphics[width=7.42 cm,angle=0]{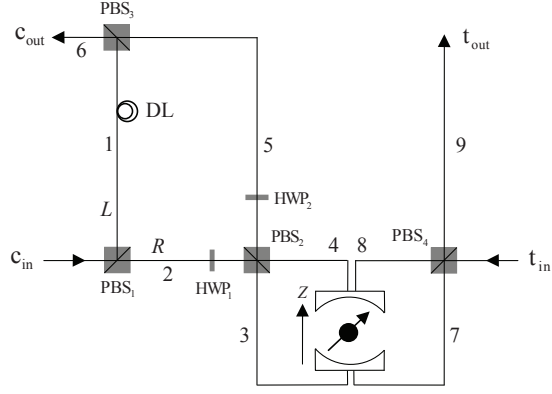}
\caption{ Quantum circuit for  implementing a deterministic photonic
two-qubit CNOT gate without additional photonic qubits via a
spin-QD-double-side-cavity system.  PBS$_i$ ($i=1,2,3,4$) transmits
the photon in the right-circularly-polarized state $\vert R\rangle$
and reflects the photon in the left-circularly-polarized state
$\vert L\rangle$, respectively. HWP$_j$ ($j=1,2$) is used to perform
a Hadamard operation on the polarization of photons, that is, $\vert
R\rangle\leftrightarrow \frac{1}{\sqrt{2}}(\vert R \rangle + \vert
L\rangle)$ and $\vert L\rangle\leftrightarrow
\frac{1}{\sqrt{2}}(\vert R \rangle - \vert L\rangle)$.  DL is the
time-delay device for making the photons from spatial modes 5 and 1
reach PBS$_3$ simultaneously, that is, fiber loops for the storage
of the photon for the time needed by the interaction between the
single photon and the QD.}
 \label{Fig2}
\end{figure}

Figure \ref{Fig2} shows the principle of the present deterministic
scheme for implementing a photonic two-qubit CNOT gate which
requires no auxiliary photonic qubits. Now, let us  describe the
quantum circuit in detail. Suppose the control photon $c$, the
target photon $t$, and the excess electron-medium $e$ in the cavity
are prepared in the states
\begin{equation}             \label{eq2}
\begin{split}
|\psi\rangle_c^p &= \alpha_c |R\rangle_c+\beta_c|L\rangle_c, \\
|\psi\rangle_t^p &= \alpha_t|R\rangle_t+\beta_t|L\rangle_t, \\
|\psi\rangle^e   &= |\downarrow\rangle,
\end{split}
\end{equation}
respectively. Here  $|\alpha_c|^2 + |\beta_c|^2 =|\alpha_t|^2 +
|\beta_t|^2=1$.

First, the control photon $c$ is sent into the input port
c$_{\text{in}}$  and it reaches PBS$_1$. PBS$_1$ transmits the
right-circularly-polarized photon and reflects the
left-circularly-polarized photon, which  results in the fact that
the control photon  in the state $|R\rangle_{c,2}$ is injected into
the cavity and the control photon  in the state $|L\rangle_{c,1}$
deserts the cavity. That is, PBS$_1$ transforms the initial state of
the whole system $|\Omega_0\rangle$  into $|\Omega_1\rangle$. Here
\begin{eqnarray}               \label{eq3}
|\Omega_0\rangle&=&|\psi\rangle_c^p\otimes|\psi\rangle_t^p\otimes|\psi\rangle^e,\\
|\Omega_1\rangle&=&(\alpha_c\alpha_t|R^\downarrow\rangle_{c,2}|R\rangle_{t}+\alpha_c\beta_t|R^\downarrow\rangle_{c,2}|L\rangle_{t}
+\beta_c\alpha_t|L\rangle_{c,1}|R\rangle_{t}+\beta_c\beta_t|L\rangle_{c,1}|L\rangle_{t})|\downarrow\rangle.
\label{eq23}
\end{eqnarray}
The subscript $i$ ($i=1,\;2,\;\cdots,$) of $|R\rangle_{c,\;i}$
($|L\rangle_{c,\;i}$) denotes the photon is in the state $|R\rangle$
$(|L\rangle)$ emitting from   spatial mode $i$. Before and after the
control photon coming from  spatial mode 2 interacts with the QD
inside the cavity, a  Hadamard operation ($H_p$, i.e., passing
through HWP$_1$ whose optical axes is set to be $22.5^\circ$) is
performed on it and a Hadamard operation ($H_e$, e.g., using a
$\pi/2$ microwave pulse or an optical pulse
\cite{manipulating1,manipulating2,manipulating3}) is also performed
on the electron, simultaneously. These two Hadamard operations
complete the transformations as following:
\begin{equation}                \label{eq5}
|R\rangle \;\;\xrightarrow{\text{$H_p$}}\;\;
\frac{1}{\sqrt2}(|R\rangle+|L\rangle), \;\;\;\;\;\;\;\;\qquad
|L\rangle \;\;\xrightarrow{\text{$H_p$}}\;\;
\frac{1}{\sqrt2}(|R\rangle-|L\rangle),
\end{equation}
\begin{equation}                \label{eq6}
|\uparrow\rangle  \;\; \xrightarrow{\text{$H_e$}}\;\;
|\rightarrow\rangle\equiv\frac{1}{\sqrt2}(|\uparrow\rangle+|\downarrow\rangle),
\;\;\;\;\;\;\;\;\qquad |\downarrow\rangle\;\;
\xrightarrow{\text{$H_e$}}\;\;
|\leftarrow\rangle\equiv\frac{1}{\sqrt2}(|\uparrow\rangle-|\downarrow\rangle).
\end{equation}
After the photon emitting from spatial mode 2 interacts with the QD
inside the cavity, one can obtain
\begin{equation}               \label{eq8}
\begin{split}
|\Omega_2\rangle=&
-\frac{1}{2}\alpha_c\alpha_t(|R^\downarrow\rangle_{c,3}+|L^\uparrow\rangle_{c,4})|R\rangle_{t}(|\uparrow\rangle
+|\downarrow\rangle) \\
&- \frac{1}{2}\alpha_c\beta_t(|R^\downarrow\rangle_{c,3}+|L^\uparrow\rangle_{c,4})|L\rangle_{t}(|\uparrow\rangle+|\downarrow\rangle) \\
&+
(\beta_c\alpha_t|L\rangle_{c,1}|R\rangle_{t}+\beta_c\beta_t|L\rangle_{c,1}|L\rangle_{t})|\leftarrow\rangle.
\end{split}
\end{equation}
PBS$_2$ transforms $|R^\downarrow\rangle_{c,3}$ and
$|L^\uparrow\rangle_{c,4}$ into $|R^\downarrow\rangle_{c,5}$ and
$|L^\uparrow\rangle_{c,5}$, respectively, and then an $H_p$ (i.e.,
passing through HWP$_2$) and an $H_e$ are performed on the control
photon and the electron, respectively. After the wave-packets
emitting from spatial modes 1 and 5 arrive PBS$_3$ simultaneously,
one can obtain
\begin{equation}                \label{eq10}
\begin{split}
|\Omega_3\rangle=&
-\alpha_c\alpha_t|R\rangle_{c,6}|R\rangle_{t}|\uparrow\rangle
-\alpha_c\beta_t|R\rangle_{c,6}|L\rangle_{t}|\uparrow\rangle+
\beta_c\alpha_t|L\rangle_{c,6}|R\rangle_{t}|\downarrow\rangle+\beta_c\beta_t|L\rangle_{c,6}|L\rangle_{t}|\downarrow\rangle.
\end{split}
\end{equation}

Next, the target photon $t$ is injected into the cavity through the
input port  t$_{\text{in}}$.  PBS$_4$ completes the transformations
$|R\rangle_{t}\xrightarrow{\text{PBS}_4}|R^\downarrow\rangle_{t,\;8}$
and
$|L\rangle_{t}\xrightarrow{\text{PBS}_4}|L^\uparrow\rangle_{t,\;7}$.
After the target photon interacts with the QD, it is emitted  from
path 9. This nonlinear interaction transforms the state of the whole
system into
\begin{equation}               \label{eq12}
\begin{split}
|\Omega_4\rangle=& (\alpha_c\alpha_t|R\rangle_{c,6}|R\rangle_{t,9}
+\alpha_c\beta_t|R\rangle_{c,6}|L\rangle_{t,9})|\uparrow\rangle +
(\beta_c\alpha_t|L\rangle_{c,6}|L\rangle_{t,9}+\beta_c\beta_t|L\rangle_{c,6}|R\rangle_{t,9})|\downarrow\rangle.
\end{split}
\end{equation}

At last, one performs an $H_e$ on the excess electron and measures
it in the  basis $\{|\uparrow\rangle, |\downarrow\rangle\}$.
According to the outcome of this measurement, a proper feed-forward
operation is performed on the control qubit to complete the CNOT
gate. If the spin is in the state $|\downarrow\rangle$,
$\sigma_z=|R\rangle\langle R|-|L\rangle\langle L|$ is performed on
the photon emitting from the output port c$_{\text{out}}$.
Otherwise, nothing is performed on the photons from both
c$_{\text{out}}$ and t$_{\text{out}}$. With these measurement and
operation, the state of the two-photon system becomes
\begin{eqnarray}               \label{13}
|\psi_{ct}\rangle = \alpha_c\alpha_t|R\rangle_{c,6}|R\rangle_{t,9}
+\alpha_c\beta_t|R\rangle_{c,6}|L\rangle_{t,9}  +
\beta_c\alpha_t|L\rangle_{c,6}|L\rangle_{t,9}+\beta_c\beta_t|L\rangle_{c,6}|R\rangle_{t,9}.
\end{eqnarray}
One can see that the state of  the target photonic qubit is flipped
when the control photonic qubit is in the state $\vert L\rangle$,
while it does not change when the control photon is in the state
$\vert R\rangle$, compared to the original state of the two-photon
system $|\psi\rangle_c^p\otimes |\psi\rangle_t^p$. That is, the
quantum circuit shown in Fig. \ref{Fig2} can be used to achieve a
deterministic CNOT gate on two photonic qubits with a success
probability of 100\% in principle.

\section{Deterministic Toffoli gate for three photonic qubits}\label{sec3}

\begin{figure}
\centering
\includegraphics[width=11.65 cm,angle=0]{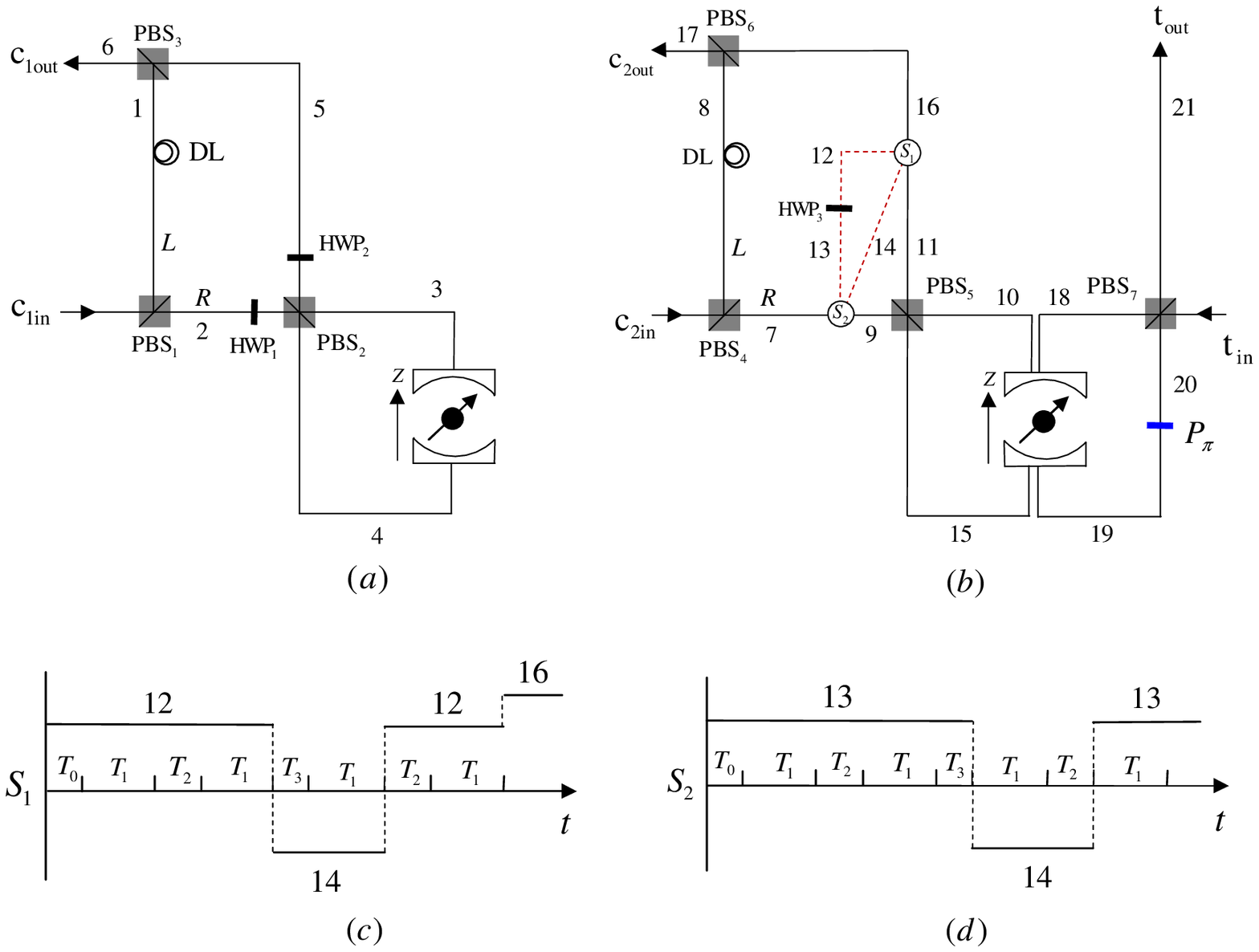}
\caption{Schematic diagram  for implementing a deterministic
photonic three-qubit Toffoli gate. The implementation of the Toffoli
gate is divided into two processes shown in (a) and (b), and the
cavity is just the same one for the first and the second control
qubits to interact with the spin in sequence. The phase shifter
P$_{\pi}$   contributes a $\pi$ phase shift to the photon passing
through it. The working states of the optical switches S$_1$ and
S$_2$ shown in (c) and (d) can be controlled accurately by computer.
S$_1$ leads the photon emitting from spatial mode 11 to 12 until
$t=T_0+2T_1+T_2$, and then leads the photon emitting from spatial
mode 11 to 14 from $t=T_0+2T_1+T_2$ to $t=T_0+3T_1+T_2+T_3$. When
$T_0+3T_1+T_2+T_3<t<T_0+4T_1+2T_2+T_3$, S$_1$ turns back to spatial
mode 12 again. $T_0$ is the time for the single-photon process
PBS$_4$-S$_2$, while $T_1$ for S$_2$-PBS$_5$-cavity-PBS$_5$-S$_1$,
$T_2$ for S$_1$-HWP$_3$-S$_2$, and $T_3$ for S$_2$-S$_1$. DL in (a)
is used for the storage of the photon for the time needed by the
interaction between the single photon and the QD, while DL in (b) is
four times of (a).} \label{Fig3}
\end{figure}

The schematic diagram for implementing a deterministic three-qubit
Toffoli gate, which performs a NOT operation on the target qubit if
and only if (iff) both the two control qubits are in the
polarization state $|L\rangle$, is shown in Fig. \ref{Fig3}. Suppose
the inputting control photons $c_1$ and $c_2$, and the target photon
$t$ are prepared in arbitrary polarization superposition states as
follows:
\begin{equation}               \label{eq14}
\begin{split}
|\psi\rangle_{c_1}^p &= \alpha_{c_1} |R\rangle_{c_1}+\beta_{c_1}|L\rangle_{c_1}, \\
|\psi\rangle_{c_2}^p &= \alpha_{c_2}|R\rangle_{c_2}+\beta_{c_2}|L\rangle_{c_2}, \\
|\psi\rangle_{t}^p &=
\alpha_{t}|R\rangle_{t}+\beta_{t}|L\rangle_{t}.
\end{split}
\end{equation}
Here $|\alpha_{c_1}|^2 + |\beta_{c_1}|^2 =|\alpha_{c_2}|^2 +
|\beta_{c_2}|^2=|\alpha_{t}|^2 + |\beta_{t}|^2=1$. To show the
principle of the present Toffoli gate explicitly, we give the
evolution of the whole system composed of two control photons  $c_1$
and $c_2$, a target photon $t$, and an excess electron-medium below.
Its initial state is
\begin{eqnarray}                  \label{eq15}
|\Xi_0\rangle=|\psi\rangle_{c_1}^p\otimes|\psi\rangle_{c_2}^p\otimes|\psi\rangle_{t}^p\otimes|\uparrow\rangle.
\end{eqnarray}
Here $|\uparrow\rangle$ is the initial state of the excess
electron-medium confined in the QD inside the cavity.

First, the first control photon $c_1$ is injected into the input
port c$_{\text{1in}}$ (depicted in Fig. \ref{Fig3}(a)). PBS$_1$
splits the control photon into two wave-packets, that is,
$|R\rangle_{c_1}\rightarrow|R^\downarrow\rangle_{c_1,2}$ and
$|L\rangle_{c_1}\rightarrow|L\rangle_{c_1,1}$. The photon  in the
state $|R^\downarrow\rangle_{c_1,2}$ is injected into the cavity,
while the photon  in the state $|L\rangle_{c_1,1}$ does not interact
with the QD inside the cavity. With the same arguments as made for
the CNOT gate above,  we find that after the first control photon
$c_1$ interacts with the QD inside the cavity, it is  emitted from
the output port c$_{1\text{out}}$ with the state of the whole system
as
\begin{equation}                     \label{eq16}
\begin{split}
|\Xi_1\rangle =&
\big[(-\alpha_{c_1}\alpha_{c_2}|R\rangle_{c_{1,6}}|R\rangle_{c_2}
-\alpha_{c_1}\beta_{c_2}|R\rangle_{c_{1,6}}|L\rangle_{c_2})|\downarrow\rangle \\
&+ (\beta_{c_1}\alpha_{c_2}|L\rangle_{c_{1,6}}|R\rangle_{c_2}
+\beta_{c_1}\beta_{c_2}|L\rangle_{c_{1,6}}|L\rangle_{c_2})|\uparrow\rangle\big]
\otimes (\alpha_{t}|R\rangle_{t}+\beta_{t}|L\rangle_{t}).
\end{split}
\end{equation}

Second, the second control photon $c_2$ is injected into the input
port $c_{2\text{in}}$ (depicted in Fig. \ref{Fig3}(b)), and then
$|R\rangle_{c_2}\xrightarrow{\text{PBS}_4}|R^\downarrow\rangle_{c_2,\;7}$
and $|L\rangle_{c_2}\xrightarrow{\text{PBS}_4}|L\rangle_{c_2,\;8}$.
The photon  in the state $|R^\downarrow\rangle_{c_2,7}$ is injected
into the cavity and interacts with the QD inside the cavity, while
the photon  in the state $|L\rangle_{c_2,8}$ does not pass through
the cavity. The interaction between the QD and the second control
photon coming from  path 7 transforms the state of the whole system
into
\begin{equation}                  \label{eq18}
\begin{split}
|\Xi_2\rangle=&
\big[(-\alpha_{c_1}\alpha_{c_2}|R\rangle_{c_{1,6}}|L^\uparrow\rangle_{c_2,11}
-\alpha_{c_1}\beta_{c_2}|R\rangle_{c_{1,6}}|L\rangle_{c_2,8})|\downarrow\rangle \\
&+
(-\beta_{c_1}\alpha_{c_2}|L\rangle_{c_{1,6}}|R^\downarrow\rangle_{c_2,11}
+\beta_{c_1}\beta_{c_2}|L\rangle_{c_{1,6}}|L\rangle_{c_2,8})|\uparrow\rangle\big]
 \otimes(\alpha_{t}|R\rangle_{t}+\beta_{t}|L\rangle_{t}).
 \end{split}
\end{equation}
Next, we lead the photon in the state $|L^\uparrow\rangle_{c_2,11}$
or $|R^\downarrow\rangle_{c_2,11}$ to spatial mode 12 with the
optical switch S$_1$, that is,
$|L^\uparrow\rangle_{c_2,11}\xrightarrow{\text{S}_2}|L^\uparrow\rangle_{c_2,12}$
and
$|R^\downarrow\rangle_{c_2,11}\xrightarrow{\text{S}_2}|R^\downarrow\rangle_{c_2,12}$.
Before the photon reaches the optical switch S$_2$, Hadamard
operations $H_p$ (i.e., passing through HWP$_3$) and $H_e$ are
performed on the photon and the electron, respectively. The
interaction between the second control photon coming from path 13
and the QD transforms the state of the system into
\begin{equation}                  \label{eq20}
\begin{split}
 |\Xi_3\rangle=&
\big[\frac{\alpha_{c_1}\alpha_{c_2}}{2}|R\rangle_{c_{1,6}}(|R^\downarrow\rangle_{c_2,11}-
|L^\uparrow\rangle_{c_2,11})(|\uparrow\rangle-|\downarrow\rangle)
-\alpha_{c_1}\beta_{c_2}|R\rangle_{c_{1,6}}|L\rangle_{c_2,8}|\leftarrow\rangle \\
&
+\frac{\beta_{c_1}\alpha_{c_2}}{2}|L\rangle_{c_{1,6}}(|R^\downarrow\rangle_{c_2,11}
+|L^\uparrow\rangle_{c_2,11})(|\uparrow\rangle-|\downarrow\rangle)
+\beta_{c_1}\beta_{c_2}|L\rangle_{c_{1,6}}|L\rangle_{c_2,8}|\rightarrow\rangle\big]\\&
\otimes(\alpha_{t}|R\rangle_{t}+\beta_{t}|L\rangle_{t}).
\end{split}
\end{equation}
Subsequently, an $H_e$ is performed on the electron, which
transforms the state of the system into
\begin{equation}                  \label{eq21}
\begin{split}
|\Xi_4\rangle=&
\big[\frac{\alpha_{c_1}\alpha_{c_2}}{\sqrt{2}}|R\rangle_{c_{1,6}}(|R^\downarrow\rangle_{c_2,11}
-|L^\uparrow\rangle_{c_2,11})|\downarrow\rangle
-\alpha_{c_1}\beta_{c_2}|R\rangle_{c_{1,6}}|L\rangle_{c_2,8}|\downarrow\rangle \\
&+
\frac{\beta_{c_1}\alpha_{c_2}}{\sqrt{2}}|L\rangle_{c_{1,6}}(|R^\downarrow\rangle_{c_2,11}
+|L^\uparrow\rangle_{c_2,11})|\downarrow\rangle +
\beta_{c_1}\beta_{c_2}|L\rangle_{c_{1,6}}|L\rangle_{c_2,8}|\uparrow\rangle\big]\\&
\otimes(\alpha_{t}|R\rangle_{t}+\beta_{t}|L\rangle_{t}).
\end{split}
\end{equation}
Next, we lead the photon in the state $|L^\uparrow\rangle_{c_2,11}$
or $|R^\downarrow\rangle_{c_2,11}$ to  spatial path 14 for
interacting  with the QD inside the cavity again by using the switch
S$_1$. The evolution of the lead-back photon can be described by the
processes $|R^\downarrow\rangle_{c_2,11}
\xrightarrow{\text{S$_1$}}|R^\downarrow\rangle_{c_2,14}
\xrightarrow{\text{S$_2$}}|R^\downarrow\rangle_{c_2,9}\xrightarrow{\text{PBS$_5$}}|R^\downarrow\rangle_{c_2,10}$
and $|L^\uparrow\rangle_{c_2,11}
\xrightarrow{\text{S$_1$}}|L^\uparrow\rangle_{c_2,14}
\xrightarrow{\text{S$_2$}}|L^\uparrow\rangle_{c_2,9}\xrightarrow{\text{PBS$_5$}}|L^\uparrow\rangle_{c_2,15}$,
respectively.

Third, as depicted in Fig. \ref{Fig3}(b), the target photon $t$ is
injected into the input port $t_{\text{in}}$ and interacts with the
QD inside the cavity after an $H_e$ is performed on the electron
described by Eq. (\ref{eq21}) and before the photon emitting from
path 11 interacts with the QD. The evolution of the target photon
$t$ can be described by the processes
$|R^\downarrow\rangle_{t}|\uparrow\rangle
\xrightarrow{\text{PBS$_7$}}|R^\downarrow\rangle_{t,18}|\uparrow\rangle
\xrightarrow{\text{cavity}}-|R^\downarrow\rangle_{t,19}|\uparrow\rangle
\xrightarrow{\text{P$_{\pi}$}}|R^\downarrow\rangle_{t,20}|\uparrow\rangle
\xrightarrow{\text{PBS$_7$}}|R^\downarrow\rangle_{t,21}|\uparrow\rangle$,
      $|R^\downarrow\rangle_{t}|\downarrow\rangle\xrightarrow{\text{PBS$_7$}}
|R^\downarrow\rangle_{t,18}|\downarrow\rangle
\xrightarrow{\text{cavity}}|L^\uparrow\rangle_{t,18}|\downarrow\rangle
\xrightarrow{\text{PBS$_7$}}|L^\uparrow\rangle_{t,21}|\downarrow\rangle$,
  $|L^\uparrow\rangle_{t}|\uparrow\rangle\xrightarrow{\text{PBS$_7$}}|L^\uparrow\rangle_{t,20}|\uparrow\rangle
\xrightarrow{\text{P$_{\pi}$}}-|L^\uparrow\rangle_{t,19}|\uparrow\rangle
\xrightarrow{\text{cavity}}|L^\uparrow\rangle_{t,18}|\uparrow\rangle
\xrightarrow{\text{PBS$_7$}}|L^\uparrow\rangle_{t,21}|\uparrow\rangle$,
and $|L^\uparrow\rangle_{t}|\downarrow\rangle
\xrightarrow{\text{PBS$_7$}}|L^\uparrow\rangle_{t,20}|\downarrow\rangle
\xrightarrow{\text{P$_{\pi}$}}-|L^\uparrow\rangle_{t,19}|\downarrow\rangle
\xrightarrow{\text{cavity}}-|R^\downarrow\rangle_{t,19}|\downarrow\rangle
\xrightarrow{\text{P$_{\pi}$}}|R^\downarrow\rangle_{t,20}\xrightarrow{\text{PBS$_7$}}|R^\downarrow\rangle_{t,21}$,
respectively.

Fourth, we perform an $H_e$ on the electron after the target photon
$t$ interacts with the QD and before the second control photon $c_2$
  in the state $|R^\downarrow\rangle_{c_2,10}$ or
$|L^\uparrow\rangle_{c_2,15}$ interacts with the QD. After
$|R^\downarrow\rangle_{c_2,10}$ and $|L^\uparrow\rangle_{c_2,15}$
 interact with the QD, the state of the whole system becomes
\begin{equation}                     \label{eq24}
\begin{split}
|\Xi_5\rangle=&
-\frac{\alpha_{c_1}\alpha_{c_2}}{2}|R\rangle_{c_{1,6}}(|R^\downarrow\rangle_{c_2,11}-|L^\uparrow\rangle_{c_2,11})
(\alpha_t|L\rangle_{t,21}+\beta_t|R\rangle_{t,21})(|\uparrow\rangle-|\downarrow\rangle) \\
&- \alpha_{c_1}\beta_{c_2}|R\rangle_{c_{1,6}}|L\rangle_{c_2,8}(\alpha_t|L\rangle_{t,21}+\beta_t|R\rangle_{t,21})|\leftarrow\rangle \\
&-
\frac{\beta_{c_1}\alpha_{c_2}}{2}|L\rangle_{c_{1,6}}(|R^\downarrow\rangle_{c_2,11}+|L^\uparrow\rangle_{c_2,11})
(\alpha_t|L\rangle_{t,21}+\beta_t|R\rangle_{t,21})(|\uparrow\rangle+|\downarrow\rangle) \\
&+
\beta_{c_1}\beta_{c_2}|L\rangle_{c_{1,6}}|L\rangle_{c_2,8}(\alpha_t|R\rangle_{t,21}+\beta_t|L\rangle_{t,21})|\rightarrow\rangle.
\end{split}
\end{equation}

Next, we lead the photon emitting from  path 11 to  path 12 by using
the switch S$_1$ ($|R^\downarrow\rangle_{c_2,11}
\xrightarrow{\text{S$_1$}}|R^\downarrow\rangle_{c_2,12}$,
$|L^\uparrow\rangle_{c_2,11}\xrightarrow{\text{S$_1$}}|L^\uparrow\rangle_{c_2,12}$)
 for passing through HWP$_3$ (i.e., an $H_p$ is performed on
$|R^\downarrow\rangle_{c_2,12}$ and $|L^\uparrow\rangle_{c_2,12}$).
Also, an $H_e$ is performed on the electron. When the photon is
emitted from spatial mode 13, it is led to the cavity by S$_2$.
After the photon interacts with the QD inside the cavity, it is
emitted from path 16 with the state of the whole system as
\begin{equation}                  \label{eq26}
\begin{split}
|\Xi_{6}\rangle=&
-\alpha_{c_1}\alpha_{c_2}|R\rangle_{c_{1,6}}|R^\downarrow\rangle_{c_2,16}(\alpha_t|L\rangle_{t,21}
+\beta_t|R\rangle_{t,21})|\downarrow\rangle \\
&- \alpha_{c_1}\beta_{c_2}|R\rangle_{c_{1,6}}|L\rangle_{c_2,8}(\alpha_t|L\rangle_{t,21}+\beta_t|R\rangle_{t,21})|\downarrow\rangle \\
&+ \beta_{c_1}\alpha_{c_2}|L\rangle_{c_{1,6}}|R^\downarrow\rangle_{c_2,16}(\alpha_t|L\rangle_{t,21}+\beta_t|R\rangle_{t,21})|\uparrow\rangle \\
&+
\beta_{c_1}\beta_{c_2}|L\rangle_{c_{1,6}}|L\rangle_{c_2,8}(\alpha_t|R\rangle_{t,21}+\beta_t|L\rangle_{t,21})|\uparrow\rangle.
\end{split}
\end{equation}
When the photons in the states $|R^\downarrow\rangle_{c_2,\;16}$ and
$|L\rangle_{c_2,\;8}$ pass through  PBS$_6$ simultaneously, the
state of the whole system becomes
\begin{equation}                  \label{eq27}
\begin{split}
|\Xi_{7}\rangle =&
-\alpha_{c_1}\alpha_{c_2}|R\rangle_{c_{1,6}}|R\rangle_{c_2,17}(\alpha_t|L\rangle_{t,21}
+\beta_t|R\rangle_{t,21})|\downarrow\rangle \\
&- \alpha_{c_1}\beta_{c_2}|R\rangle_{c_{1,6}}|L\rangle_{c_2,17}(\alpha_t|L\rangle_{t,21}+\beta_t|R\rangle_{t,21})|\downarrow\rangle \\
&+ \beta_{c_1}\alpha_{c_2}|L\rangle_{c_{1,6}}|R\rangle_{c_2,17}(\alpha_t|L\rangle_{t,21}+\beta_t|R\rangle_{t,21})|\uparrow\rangle \\
&+
\beta_{c_1}\beta_{c_2}|L\rangle_{c_{1,6}}|L\rangle_{c_2,17}(\alpha_t|R\rangle_{t,21}+\beta_t|L\rangle_{t,21})|\uparrow\rangle.
\end{split}
\end{equation}

Fifth, an $H_e$ is first performed on the electron and then a
measurement in the basis $\{|\uparrow\rangle,|\downarrow\rangle\}$
is taken on it. In order to complete the Toffoli gate with a success
probability of 100\%, $-\sigma_z=-|R\rangle\langle R| +
|L\rangle\langle L|$ and $\sigma_x =|R\rangle\langle L| +
|L\rangle\langle R|$  are performed on the outing control photon
$c_1$ and the outing target photon $t$, respectively iff the spin is
in $|\uparrow\rangle$. $\sigma_x $ is performed on the outing target
photon $t$ iff the spin is in $|\downarrow\rangle$. After these
operations, the finial state of the three-photon system becomes
\begin{equation}                  \label{28}
\begin{split}
|\Psi_T\rangle =&  \alpha_{c_1}\alpha_{c_2}|R\rangle_{c_{1,6}}|R\rangle_{c_2,17}(\alpha_t|R\rangle_{t,21}+\beta_t|L\rangle_{t,21})  \\
&+ \alpha_{c_1}\beta_{c_2}|R\rangle_{c_{1,6}}|L\rangle_{c_2,17}(\alpha_t|R\rangle_{t,21}+\beta_t|L\rangle_{t,21})  \\
&+ \beta_{c_1}\alpha_{c_2}|L\rangle_{c_{1,6}}|R\rangle_{c_2,17}(\alpha_t|R\rangle_{t,21}+\beta_t|L\rangle_{t,21})  \\
&+
\beta_{c_1}\beta_{c_2}|L\rangle_{c_{1,6}}|L\rangle_{c_2,17}(\alpha_t|L\rangle_{t,21}+\beta_t|R\rangle_{t,21}).
\end{split}
\end{equation}
One can see that the state of the target photonic qubit $t$ is
flipped when both the two control photonic qubits $c_1$ and $c_2$
are in the state $\vert L\rangle$, compared to the initial state of
the photon system composed of the three photonic qubits. That is,
the quantum circuit shown in Fig. \ref{Fig3} can be used to
construct a Toffoli gate on a three-photon system in a deterministic
way.

\section{Discussion and summary}\label{sec4}

Let us discuss the feasibility to implement our proposals in a
promising system with a GaAs- or InAs-based QD confined in an
optical resonant microcavity with two partially reflective mirrors.

We can  briefly review the reflection and the transmission
coefficients of the cavities, for describing the photon-spin
interaction in the system. It is known that they can be obtained by
solving the Heisenberg equations of motion for the cavity field
operator $\hat{a}$ and the QD dipole operator $\sigma_-$
\begin{equation}               \label{eq29}
\begin{split} \frac{d \hat{a}}{d t} =& -\left[
i(\omega_c-\omega)+\kappa+\frac{\kappa_s}{2}
\right]\hat{a}-g\sigma_-
-\sqrt{\kappa}\,\hat{a}_{in}-\sqrt{\kappa}\,\hat{a}_{in}'+\hat{H}, \\
\frac{d \sigma_-}{d t} &= -\left[
i(\omega_{X^-}-\omega)+\frac{\gamma}{2}
\right]\sigma_--g\sigma_z\,\hat{a}+\hat{G},
\end{split}
\end{equation}
combining with the input-output relations for the cavity
\cite{Heisenberg}
\begin{eqnarray}               \label{eq30}
\hat{a}_r=\hat{a}_{in}+\sqrt{\kappa}\,\hat{a},\;\;\;\;\;\;\;\;\;\;\;\;\;\;\;\;
\hat{a}_t=\hat{a}_{in}'+\sqrt{\kappa}\,\hat{a},
\end{eqnarray}
 and taking
$\langle\sigma_z\rangle\approx-1$. Hu \emph{et al.} \cite{Hu2}
presented an explicit expression for the transmission coefficient
$t(\omega)$ and the reflection coefficient $r(\omega)$. They can be
expressed as
\begin{eqnarray}               \label{eq31}
t(\omega)=\frac{-\kappa\left[i(\omega_{X^{-}}-\omega)+\frac{\gamma}{2}\right]}{\left[i(\omega_{X^{-}}\omega)
+\frac{\gamma}{2}\right]\left[i(\omega_c-\omega)+\kappa+\frac{\kappa_s}{2}\right]+g^2},\;\;\;\;\;\;\;\;\;\;\;\;\;\;\;\;
r(\omega)=1+t(\omega).
\end{eqnarray}
Here $\omega$, $\omega_c$, and $\omega_{X^-}$ denote the frequencies
of the external field (probe photon), the cavity mode, and the $X^-$
transition, respectively. $g$ denotes the $X^-$ cavity coupling
rate. $\gamma/2$, $\kappa$, and $\kappa_s/2$ denote the decay rates
of the $X^-$ dipole, the cavity field, and the leaky modes (side
leakage), respectively. As shown in Fig. \ref{Fig1}(a), $\hat{H}$
and $\hat{G}$ are the noise operators related to the reservoirs
needed to conserve the commutation. $\hat{a}_{in}$ ($\hat{a}_{in}'$)
 and $\hat{a}_r$ and ($\hat{a}_t$) are the input field operators and the output field operators, respectively.

The reflected and the transmitting lights feel phase shifts due to
the complex reflection and transmission coefficients given by Eq.
(\ref{eq31}), and the phase shift can be adjusted by
$\omega-\omega_c$ ($\omega_c=\omega_{X^-}$). In our work, we set the
cavity to be resonant with the $X^-$ and the input photons, that is,
$\omega_c=\omega_{X^-}=\omega$, and the reflection and the
transmission coefficients of the coupled (hot, that is, $g\neq0$)
cavity described by  Eq. (\ref{eq31}) can be simplified as
\begin{eqnarray}              \label{eq32}
\;\;\;\;\;\;\;\;\;\;\;\;\;\;\;\; r=1+t,
\;\;\;\;\;\;\;\;\;\;\;\;\;\;\;\;\;\;\;\;\;\;\;\;\;\;\;\;
t=-\frac{\kappa\frac{\gamma}{2}}{\frac{\gamma}{2}\left[\kappa+\frac{\kappa_s}{2}\right]+g^2}.
\end{eqnarray}
By taking $g$=0, the reflection and the transmission coefficients of
the uncoupled cavity (cold cavity) can be written as
\begin{eqnarray}              \label{eq33}
r_0=\frac{\frac{\kappa_s}{2}}{\kappa+\frac{\kappa_s}{2}},\;\;\;\;\;\;\;\;\;\;\;\;\;\;\;\;\;\;\;\;\;\;t_0=-\frac{\kappa}{\kappa+\frac{\kappa_s}{2}}.
\end{eqnarray}
Therefore, in a realistic spin-QD-double-side-cavity unit, the
change of the input photon states in the system can be described as
\cite{Appli1}
\begin{equation}              \label{eq34}
\begin{split}
|R^\uparrow
\uparrow\rangle&\rightarrow|r||L^\downarrow\uparrow\rangle+|t||R^\uparrow\uparrow\rangle,\;\;\;\;\;\;\;\;\quad\quad\quad
|L^\downarrow\uparrow\rangle \rightarrow|r||R^\uparrow \uparrow\rangle  +|t||L^\downarrow \uparrow\rangle, \\
|R^\downarrow\downarrow\rangle
&\rightarrow|r||L^\uparrow\downarrow\rangle+|t||R^\downarrow
\downarrow\rangle,\;\;\;\;\;\;\;\;\quad\quad\quad
|L^\uparrow\downarrow\rangle\rightarrow|r||R^\downarrow\downarrow\rangle+|t||L^\uparrow \downarrow\rangle, \\
|R^\downarrow\uparrow\rangle&\rightarrow-|t_0||R^\downarrow\uparrow\rangle-|r_0||L^\uparrow\uparrow\rangle,\;\;\;\;\;\;\;\;\quad\;\;
|L^\uparrow\uparrow\rangle\rightarrow-|t_0||L^\uparrow\uparrow\rangle-|r_0||R^\downarrow \uparrow\rangle, \\
|R^\uparrow\downarrow\rangle&\rightarrow
-|t_0||R^\uparrow\downarrow\rangle-|r_0||L^\downarrow
\downarrow\rangle,\;\;\;\;\;\;\;\;\quad\;\
|L^\downarrow\downarrow\rangle
\rightarrow-|t_0||L^\downarrow\downarrow\rangle-|r_0||R^\uparrow\downarrow\rangle.
\end{split}
\end{equation}

First, we consider an ideal case in which the side leakage
$\kappa_s$ is much lower than the output coupling rate $\kappa$
(that is, $\kappa_s$ is negligible). Hu \emph{et al.} \cite{Hu2}
pointed out that under this condition, $|t_0(\omega)|\rightarrow 1$,
$|r_0(\omega)|\rightarrow 0$, $|t(\omega)|\rightarrow 0$, and
$|r(\omega)|\rightarrow 1$ can be achieved in the strong coupling
regime $g>(\kappa,\gamma)$. In this time, Eq. (\ref{eq34}) becomes
Eq. (\ref{eq1}), and the fidelities and the efficiencies  of our
gates can achieve unity. Unfortunately, this is a big challenge for
QD-micropillar cavities although a significant progress has been
made \cite{challenge,observed1}. If the cavity side leakage
$\kappa_s$ which will cause bit-flip errors is taken into account,
the fidelities of our CNOT and Toffoli gates  can be calculated as
\begin{eqnarray}             \label{eq35}
F_{\text{CT}}=\left[\frac{|t_0|+|r|}{2}\right]^2,\;\;\;\;\;\;\;\;\;\;\;\;\;\;\;\;\;\;\;\;
F_{\text{T}}=\left[\frac{\xi_1+2\xi_2-\xi_3}{32}\right]^2,
\end{eqnarray}
where
\begin{equation}             \label{eq36}
\begin{split}
\xi_1=& (|t_0|-|r_0|-|t|+|r|) \times [|r_0|(|t_0|-|r_0|)(|t_0|-|r_0|+|r|-|t|)^2 \\
       &+ |r_0|(|r|-|t|)(|t_0|-|r_0|-|r|+|t|)^2+4|t_0|(|r|-|t|)+4(|t_0|-|r_0|)], \\
\xi_2= &|r|(|t_0|-|r_0|)(|t_0|-|r_0|-|r|+|t|)^2+|r|(|r|-|t|)(|t_0|-|r_0|+|r|-|t|)^2 \\
       &+ 4|t|(|t_0|-|r_0|)+4(|r|-|t|), \\
\xi_3=& |r_0|(|t_0|-|r_0|+|r|-|t|)^2(|t_0|-|r_0|-|r|+|t|)^2.
\end{split}
\end{equation}
Here $F=|\langle\Psi_r|\Psi_i\rangle|^2$. $\vert \Psi_r\rangle$
represents the finial state in a realistic QD-cavity system in our
protocols, whereas $\vert \Psi_i\rangle$ represents the final state
in the ideal condition.

In our work, the efficiency of the gate is defined as the ratio of
the the number of the outputting photons to the inputting photons.
 The efficiencies of the CNOT and Toffoli gates  can be
expressed as
\begin{eqnarray}             \label{eq37}
\eta_{\text{CT}}=\frac{1}{3}\left[\frac{1}{2}+\frac{5\zeta}{4}\right],\;\;\;\;\;\;\;\;\;\;\;\;\;\;\;\;\;\;\;\;\;\;\;
\eta_{\text{T}}=\frac{1}{4}\left[1+\frac{5\zeta}{4}+\frac{\zeta^4}{32}\right].
\end{eqnarray}
Here $\zeta=|t_0|^2+|r_0|^2+|t|^2+|r|^2$.

\begin{figure}
\centering
\includegraphics[width=6 cm,angle=0]{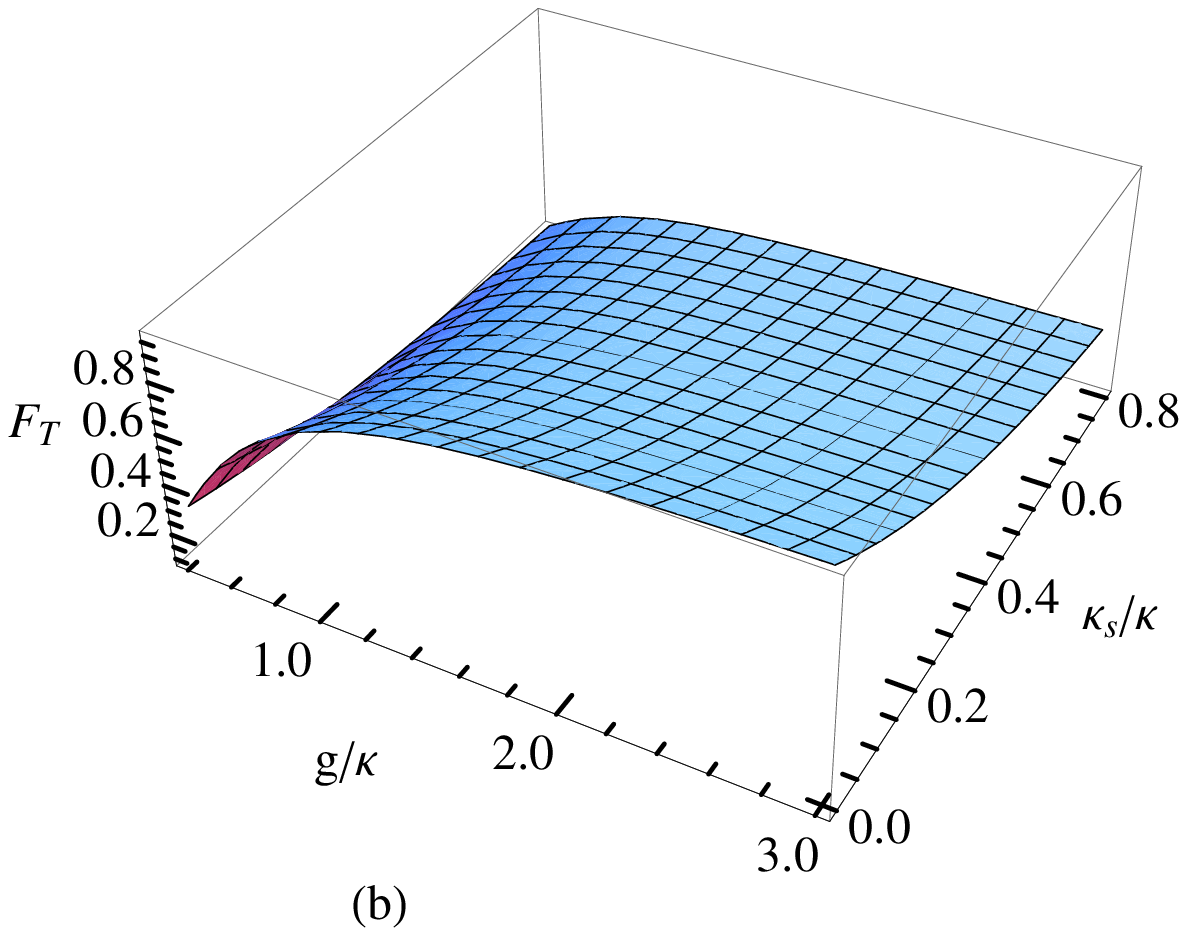}
\;\;\;\;\;\;\;\;\;
\includegraphics[width=6 cm,angle=0]{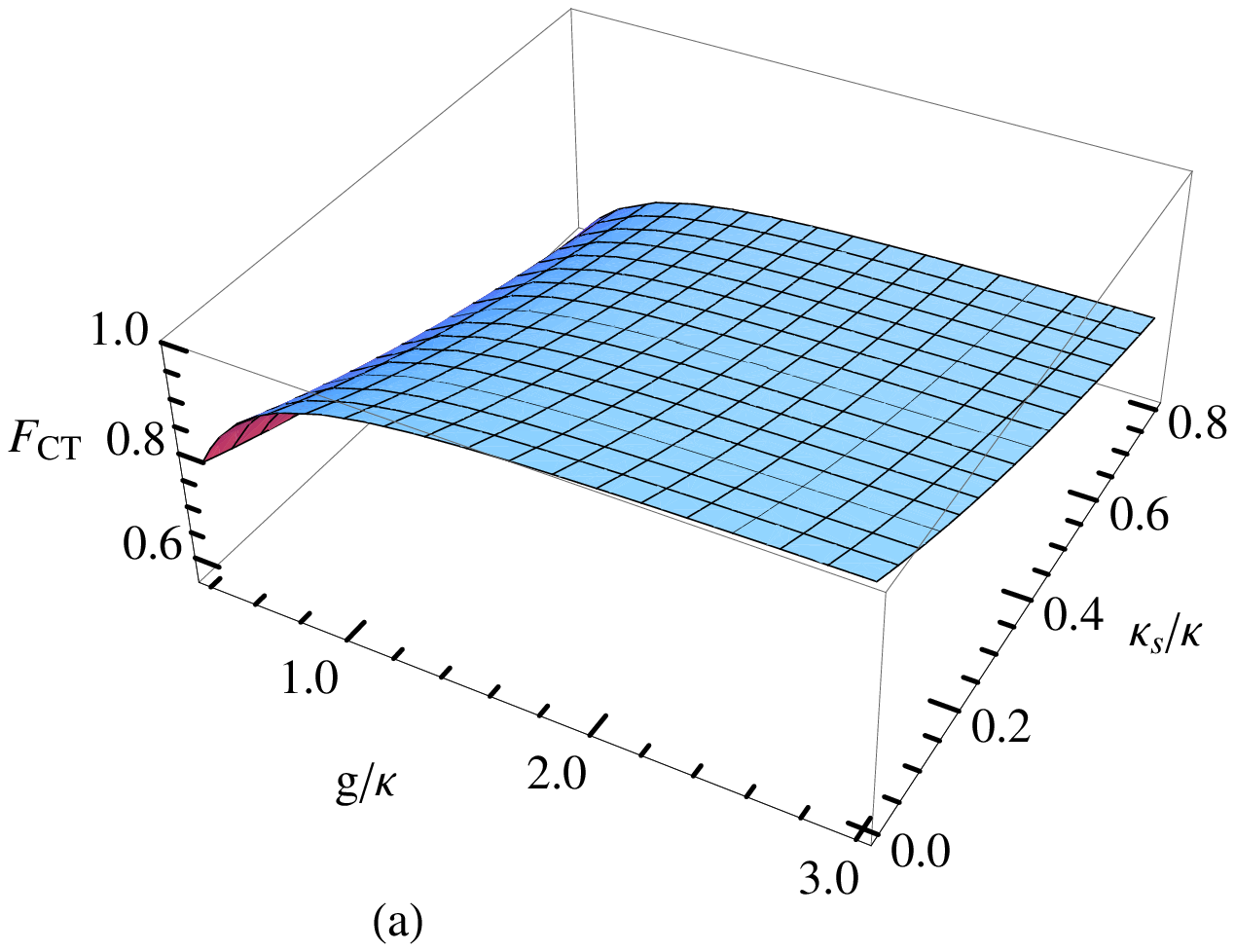}
\caption{ The fidelities of the present quantum gates as a function
of the coupling strength $g/\kappa$ and the side leakage rate
$\kappa_s/\kappa$. (a) The fidelity of the CNOT gate $F_{CT}$;  (b)
The fidelity of the Toffoli gate $F_{T}$.} \label{Fig4}
\end{figure}

\begin{figure}          
\centering
\includegraphics[width=6 cm,angle=0]{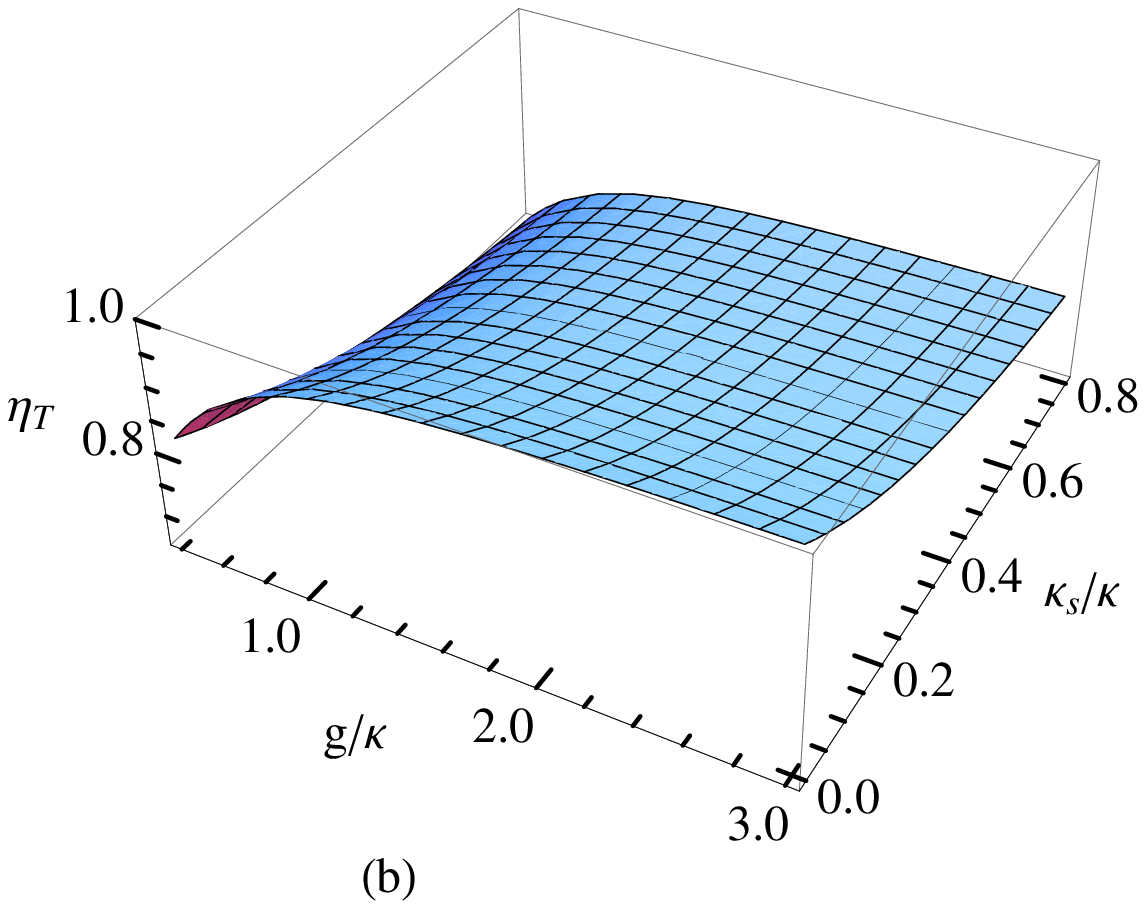}
\;\;\;\;\;\;\;\;\;
\includegraphics[width=6 cm,angle=0]{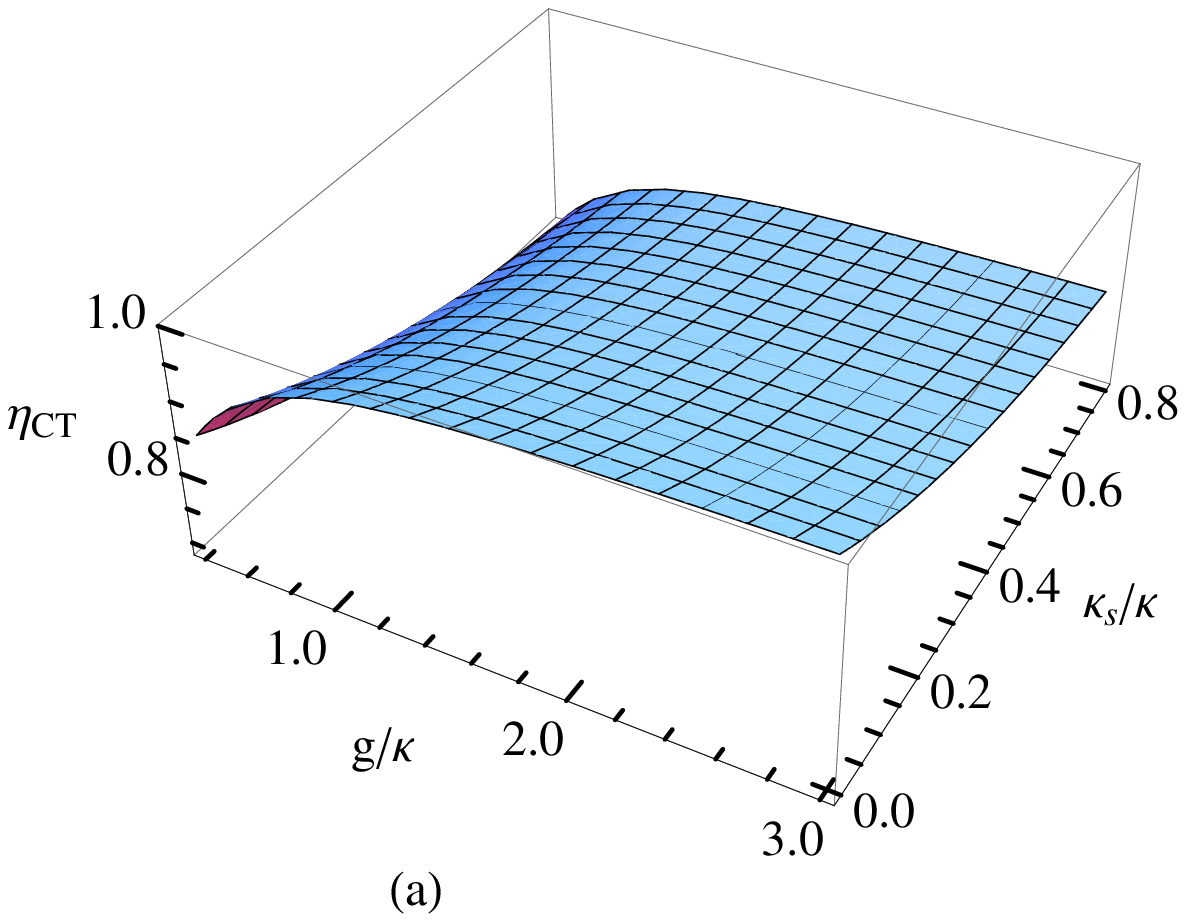}
\caption{  The efficiencies of the present deterministic
photon-qubit gates as a function of $g/\kappa$ and
$\kappa_s/\kappa$. (a) The efficiency of the CNOT gate $\eta_{CT}$;
(b) The efficiency of the Toffoli gate $\eta_{T}$.} \label{Fig5}
\end{figure}

For QD-cavity systems, strong coupling has been demonstrated.
Forchel group \cite{challenge} observed $g/(\kappa+\kappa_s)
\approx0.5$ for a diameter $d\approx 1.5\; \mu m$ micropillar with
the quality factor $Q=4\times8800$ In(Ga)As QD-microcavity.
$g/(\kappa+\kappa_s)\approx2.4$ for $d\approx 1.5\; \mu m$, $g=80\;
\mu eV$, and $Q=4\times10^4$ In(Ga)As QD-microcavity is also
observed by this group \cite{challenge,observed1}. Yoshie
\cite{observed2} reported $g/\kappa=2.4$ for QD-nanocavity. Hu
\emph{et al.} \cite{Hu4} reported $g/(\kappa+\kappa_s)\approx1.0$
for In(Ga)As QD-microcavity in 2011. In 2011, Young \emph{et al.}
\cite{ABYoung} confirmed $g>(\kappa+\kappa_s+\gamma)/4$ in pillar
microcavity. Fig. \ref{Fig4} and Fig. \ref{Fig5} present the
fidelities and the efficiencies of our gates vary with $\kappa_s
/\kappa$ and $g /\kappa$. They indicate that the fidelities and the
efficiencies decrease with $\kappa_s/\kappa$, and increase with $g
/\kappa$. Let us now present some data in detail. When
$g/\kappa=2.4$ and $\kappa_s/\kappa=0.5$, $F_{\text{CNOT}}=80.3\%$,
$F_{\text{T}}=48.4\%$ with  $\eta_{\text{CNOT}}=86\%$ and
$\eta_{\text{T}}=82.9\%$. If $\kappa_s\ll\kappa$ and $g/\kappa=2.4$,
$F_{\text{CNOT}}=99.1\%$,
 $F_{\text{T}}=95.8\%$ with $\eta_{\text{CNOT}}=99.3\%$ and
$\eta_{\text{T}}=99.05\%$.  From Fig. \ref{Fig4} and Fig.
\ref{Fig5}, we find that by taking a low $\kappa_s/\kappa$, a
near-unity fidelity and efficiency can be achieved in the
strong-coupling regime. Moreover, a high fidelity and a high
efficiency are achieved possibly in the weak-coupling regime.

Hu \emph{et al.} \cite{Hu2,Hu4} showed that besides the side leakage
of the cavity, the electron-spin decoherence, the exciton dephasing,
and the imperfect optical selection rule reduce the fidelity of a
gate by few percents, respectively. The electron-spin decoherence
decreases the fidelities of the gates by a factor
\begin{eqnarray}             \label{eq35}
\left[1+\exp(-\Delta t /T_2^e)\right]/2.
\end{eqnarray}
Here $T_2^e$ and $\Delta t$ are the electron spin coherence time and
the time interval between two input photons encoded for the gates,
respectively. The electron-spin decoherence can reduce the
fidelities by few presents as $T_2^e$ could be maintained for more
than $3\;\mu$s using spin-echo techniques
\cite{cohertime1,cohertime2} and $\Delta t$, which limited by the
critical photon and the cavity photon life time and should be
shorter than $T_2^e$ for getting a high fidelity, can be longer than
ns \cite{interval}. The exciton dephasing caused by the exciton
decoherence in a QD reduces the fidelities by a factor
\begin{eqnarray}             \label{eq35}
\left[1-\exp(-\tau/T_2)\right].
\end{eqnarray}
$\tau$ and $T_2$ are the cavity photon lifetime and the exciton
coherence time, respectively. The trion phasing induces the state of
the electron with equal polarized superposition
$(|\uparrow\rangle|+|\downarrow\rangle)/\sqrt{2}$ to be
\begin{eqnarray}
\rho^e(t)=\frac{1}{2} \left(\begin{array}{cc}
1&  e^{-t/2T_2}\\
e^{-t/2T_2} & 1
\end{array}\right),
\end{eqnarray}
as the information of the photonic qubits is transformed into the
electron through the excitonic state. It has been reported that the
exciton slightly decreases the fidelities because the exciton
dephasing includes the optical dephasing and the spin dephasing. In
self-assembled In(Ga) QD, the optical coherence time of exciton can
be maintained several hundreds of picoseconds
\cite{opticldeph1,opticldeph2,opticldeph3} and it is ten times
longer than the cavity photon lifetime ($\tau\sim$ tens of
picoseconds). The trion coherence time of the exciton ($T_2>100$ ns
has been reported) is at least three orders of longer than the
cavity photon lifetime \cite{spindeph1,spindeph2,spindeph3}. The
imperfect optical selection rule caused by the heavy-light hole
mixing \cite{opticselec} in a realistic QD could be reduced by
engineering the shape and the size of QDs or by  choosing different
types of QDs.

The time interval between the inputting photons $\Delta t=\tau/n_0$.
Here $n_0=\gamma^2/2g^2$ \cite{critical} is the number of the
critical photons and $\tau$ is the cavity photon lifetime.
$n_0=2\times10^{-3}$ and $\tau\sim10$ ns can be achieved
\cite{interval} for a micropillar microcavity with  $d=1.5$ $\mu$m
and $Q=1.7\times$ 10$^4$. Therefore,  the time difference between
the inputting photons can be longer than $\tau/n_0$=4.5 ns
\cite{interval} and it is several orders shorter than the single
electron charged QD spin coherence time  $\sim\mu$s
\cite{cohertime1,cohertime2,cohertime3,cohertime4,cohertime5,cohertime6}.
As the CNOT and Toffoli gates are encoded on two and three photons,
respectively,   the speed of the photon-qubit for those two gates
interacting with the spin should be less than $\tau/2$ and $\tau/3$,
respectively.

Schemes for realizing the quantum gate on photonic qubits by means
of the light-matter interactions have been received much attention.
An auxiliary photonic qubit or qudit, as employed in
\cite{kerr1,kerr2,Toffoli-Kerr,Ionicioiu}, is unnecessary in our
schemes. Our scheme as shown in Fig. \ref{Fig2}, 2 two-qubit
entangling gates acting on light-matter systems are required, less
than the ones in \cite{Yang,Duan} which require 3 hybrid gates.
Refs. \cite{Hu2,Devitt,Feng} discussed the photonic entanglement
assisted by an atom or a QD. However, based on parity-check gates,
an additional photonic qubit is necessary to construct a two-qubit
quantum gate. Procedure for multi-qubit gates using parity-check
gates is an open question. The circuits of the photonic qubit gates
assisted by matter-medium are mostly focused on  two-qubit cases
\cite{kerr1,Yang,Duan,Wong}, while the ones for multi-qubit systems
are much more complex. Moreover, it is usual not an appealing method
to realize a multi-qubit gates by means of two-qubit gates. For
example, the synthesis of a Toffoli gate requires 5 two-qubit
entangling gates assisted by a qudit \cite{Ionicioiu}, or requires 6
CNOT gates and some single-qubit gates \cite{optimal}. In our work,
we  not only investigated the realization of a photonic two-qubit
CNOT gate, but also generalized it to the three-qubit case.
Moreover, our schemes are based on a QD-cavity system. It is easier
to trap a QD inside the cavity than that for an atom, and the speed
of the optical coherent manipulation of a QD is far faster than an
atom. Compared with the gates based on spin-QD-single-side-cavity
systems, the ones based on spin-QD-doubled-side-cavity systems are
more robust and flexible \cite{Hu2}.

In summary, we have investigated the possibility of achieving
scalable photonic quantum computing by the giant optical circular
birefringence induced by a singly charged QD spin in a double-sided
optical microcavity as a result of cavity QED and have proposed an
attractive scheme for a deterministic CNOT gate on two photonic
qubits by two single-photon input-output processes and the readout
on an electron-medium spin. It requires no additional photonic
qubits, different from those based on cross-Kerr nonlinearity or
parity-check gates. Moreover, we have presented a deterministic
scheme for implementing a three-qubit Toffoli gate on photon
systems. In our schemes, the spin-QD-double-side-cavity system is
only a solid medium. When the ratio of the side leakage to the
cavity loss is low, a near-unity fidelity can be achieved in the
strong-coupling regime and a high fidelity can be achieved in the
weak-coupling regime. With these two quantum gates on photonic
qubits and single-photon unitary operations, universal quantum
computing can be achieved  on photon systems in principle.

\section*{ACKNOWLEDGEMENTS}

This work is supported by the National Natural Science Foundation of
China under Grant No. 11174039,  NECT-11-0031, and the Fundamental
Research Funds for the Central Universities.


\begin{thebibliography}{99}
\bibitem{interval} By taking $g/(\kappa+\kappa_s)=1.0$,$\kappa_s/\kappa=0.7$ and $\gamma=0.1\kappa$ for a micropillar
microcavitr with diameter $d=1.5 \mu$ m, $Q=1.7\times10^4$, one can
get $n_0=2\times10^{-3}$, $\tau=9$ ps, and  $\tau/n_0=4.5 $ ns.
\end{thebibliography}
\end{document}